\documentclass[osajnl,twocolumn,showpacs,superscriptaddress,10pt]{revtex4-1} 
\usepackage{amsmath,amssymb,graphicx,comment}

\def\U#1{{%
\def\O{\mbox{O}}
\def\u{\mbox{u}}
\mathcode`\u=\mu
\mathcode`\O=\Omega
\mathrm{#1}}}

\def\ii{{\mathrm{i}}}

\def\sub#1{_{\scriptsize\mbox{#1}}}

\def\degree{\mbox{$^\circ$}}
\def\vct#1{{\mathchoice{\mbox{\boldmath$#1$}}{\mbox{\boldmath$#1$}}%
  {\mbox{\scriptsize\boldmath$#1$}}{\mbox{\scriptsize\boldmath$#1$}}}}

\begin{document}

\title{Brewster effect in metafilms composed of bi-anisotropic
split-ring resonators}

\author{Yasuhiro Tamayama}\email{Corresponding author: tamayama@vos.nagaokaut.ac.jp}
\affiliation{Department of Electrical Engineering, Nagaoka University of
Technology, 1603-1 Kamitomioka, Nagaoka, Niigata 940-2188, Japan}

\begin{abstract}
The Brewster effect is extended to single-layer metafilms. In contrast 
to bulk media, the Brewster effect in metafilms can be realized by
tailoring the radiation pattern of a distribution of meta-atoms rather
than the effective medium parameters. A metafilm composed of
bi-anisotropic split-ring resonators is designed based on the theory,
and its characteristics are numerically analyzed. The simulation
demonstrates that there exists a condition for which 
the polarization of the reflected wave
becomes independent of the incident polarization at a particular angle
of incidence. 
\end{abstract}


\maketitle 


The Brewster effect arises in connection with the laws of reflection and refraction of
electromagnetic waves at an interface between two different
media. 
When an electromagnetic wave is incident on the interface at a particular
angle of incidence, which is called the Brewster angle, the polarization
of the reflected wave becomes independent of the incident polarization. 
It can alternatively be considered an 
antireflection phenomenon because the
reflected wave vanishes for a certain polarization of the incident wave. 
This effect is applied, for example, to generate polarized light
from unpolarized light and to suppress reflection losses at intracavity
elements.  

The antireflection condition is only satisfied for transverse-magnetic
(TM) waves (horizontally polarized waves) 
in naturally occurring media because such materials do not exhibit a
magnetic response in high frequency regions, i.e., microwave, terahertz,
and optical regions.  
It was not until
metamaterials~\cite{smith04_sci,wang_b09_jopt,tao11,soukoulis11,zheludev12}
were developed that the antireflection effect was experimentally 
observed for non-TM polarizations~\cite{tamayama06,watanabe08}. As far, 
the Brewster effect has been studied in magnetic
media~\cite{doyle80,sastry87,futterman95,fu05},
anisotropic media~\cite{grzegorczyk05,tanaka06,shen06,shu07}, chiral
(bi-isotropic) media~\cite{bassiri88,lakhtakia89,tamayama08}, 
bi-anisotropic media~\cite{lakhtakia92}, and plasmonic
media~\cite{alu11,akozbek12,argyropoulos12}.

The physical meaning of the Brewster
effect is based on
the radiation patterns of the electric dipole moment and magnetic moment
induced in the material~\cite{sastry87}. 
For simplicity, a plane electromagnetic wave is assumed to be 
incident from a vacuum onto a
pure dielectric. 
When a transverse-electric (TE) 
wave (vertically polarized wave) 
is incident on the medium, the
direction of vibration of the induced electric dipoles is
perpendicular to the plane of incidence. The radiation pattern 
in the incident plane 
is isotropic, so that the reflected wave does not
vanish for any angle of incidence. 
On the other hand, for horizontal incident polarization, 
the direction of oscillation of the electric dipoles is in the incident plane
and the radiation pattern in the incident plane 
has a null in the direction of vibration. 
If the propagation direction of the reflected wave coincides with this
null direction, the horizontally polarized reflected wave vanishes. 
Therefore, the reflected wave becomes vertically polarized regardless of
the incident polarization for this propagation direction. 

Although the Brewster effect is thought of as a phenomenon that occurs at 
an interface between two
bulk media, the principle can be extended to 
single-layer metafilms. 
In this study, it is shown that
the Brewster effect, in which the polarization of the reflected wave
becomes independent of the incident polarization at a particular angle
of incidence~\cite{bassiri88,lakhtakia89,lakhtakia92,tamayama08}, can be
achieved in metafilms by tailoring the
directions of the electric dipole moment and magnetic moment in the
constituent meta-atoms and by suitably arranging 
the spatial distribution of those meta-atoms.
The method is developed
through a theoretical analysis of the polarization of the radiated wave
from an array of bi-anisotropic
split-ring resonators
(SRRs)~\cite{katsarakis04,linden04,rockstuhl06,tao09}. 
The reflection and transmission characteristics of the metafilm
are analyzed using a
finite-difference time-domain (FDTD) method. 
The results demonstrate that there exists a condition for which the
reflected wave is vertically polarized
independent of the incident polarization.


\begin{figure}[tb]
\begin{center}
\includegraphics[scale=0.55]{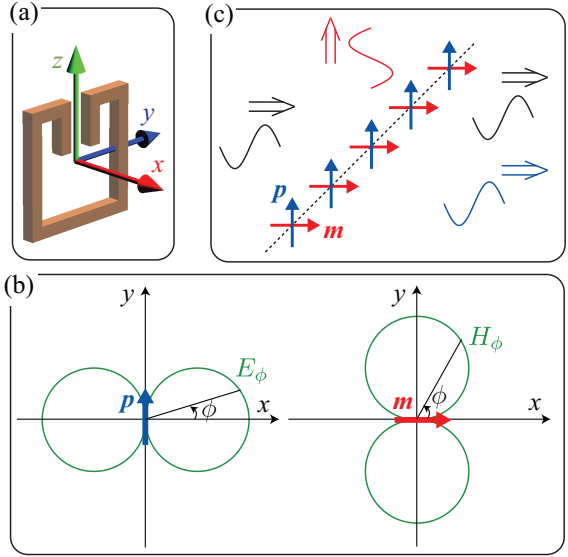}
\caption{(a) Geometry of the bi-anisotropic SRR and
of the coordinate system. (b) Radiation patterns in the $xy$ plane 
of $\vct{p}$ and $\vct{m}$ induced in
 the SRR. (c) Example of the relationship among the 
 SRRs and the propagation directions of the incident,
 reflected, and transmitted waves to achieve the Brewster effect in the
 metafilm. The incident wave is shown in black
 and the radiation from $\vct{p}$ ($\vct{m}$) is shown in blue (red).}
\label{fig:theory}
\end{center}
\end{figure}

To develop a method for achieving the Brewster effect in metafilms
of bi-anisotropic
SRRs, consider the radiation pattern of the single
SRR in Fig.\,\ref{fig:theory}(a). 
The electric dipole moment $\vct{p}$ is in the $y$ direction and the
magnetic moment $\vct{m}$ is in the $x$ direction, assuming that 
the incident electromagnetic waves only excite
circular currents in the SRR. 
The radiation patterns in the $xy$ plane of $\vct{p}$ and $\vct{m}$ are
sketched in Fig.\,\ref{fig:theory}(b). The radiation pattern 
of $\vct{p}$ ($\vct{m}$) has 
maxima (nulls) at $\phi = 0\degree$ and $180\degree$ 
and nulls (maxima) at $\phi = 90\degree$ and $270\degree$.
The polarization direction 
of the wave radiated from $\vct{p}$ is in the $xy$
plane (horizontally polarized) and that from $\vct{m}$ is perpendicular
to the $xy$ plane (vertically polarized). 
Thus, the polarization of the radiated wave in the directions of 
$\phi = 0\degree$ and $180\degree$ ($\phi = 90\degree$ and 
$270\degree$) is always horizontal (vertical).

The radiation pattern of the metafilm is determined by the product of
the radiation pattern of the constituent meta-atom and the array
factor~\cite{johnson92},
which is determined by the phase
difference between waves radiated from neighboring meta-atoms. 
If the array factor vanishes
for all angles except $\phi = 0\degree$ and $180\degree$ ($90\degree$ and
$270\degree$), the radiated wave is horizontally (vertically)
polarized. 
Therefore, when the propagation
direction of the reflected wave is equal to 
$\phi = 0\degree$ or $180\degree$ ($90\degree$ or $270\degree$)
as a result of tailoring the configuration of the SRRs and the propagation
direction of the incident wave, the reflected wave
becomes horizontally (vertically) polarized independent of the incident
polarization. 

To realize this Brewster condition, 
the SRRs should be arranged, for example,  
periodically in the $z$ direction and in the direction 
$\phi = 45\degree$, as depicted in Fig.\,\ref{fig:theory}(c). 
When the electromagnetic wave is incident from 
$\phi = 180\degree$, the metafilm radiates electromagnetic waves
in the directions $\phi = 0\degree$ and
$90\degree$, as determined by the phase difference between 
radiation from neighboring SRRs (i.e., conservation of linear momentum in the
direction parallel to the surface of the metafilm). 
That is, the array factor for this case vanishes for all angles except
$\phi = 0 \degree$ and $90\degree$.
If the incident wave is horizontally polarized, the metafilm is
excited and a vertically polarized wave is reflected in the direction
$\phi = 90\degree$, as indicated by the red wave in Fig.\,\ref{fig:theory}(c). 
On the other hand, for vertical incident polarization, 
the metafilm is not excited and the reflection vanishes. 
Thus, the reflected wave is vertically polarized independent of the
incident polarization. 
(When the electromagnetic wave is incident from the direction 
$\phi = 90\degree$, 
the reflected wave is horizontally
polarized independent of the incident polarization
according to the above theory. However, that does not occur
for a metafilm composed of actual bi-anisotropic SRRs, as explained later.)


\begin{figure}[tb]
\begin{center}
\includegraphics[scale=0.7]{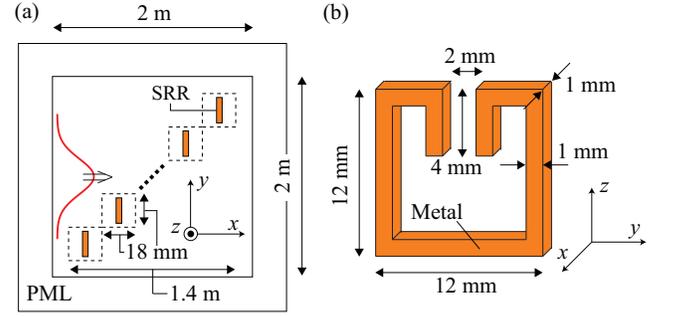}
\caption{(a) Schematic of the simulation system. 
(b) Geometrical parameters of the bi-anisotropic SRR. 
The unit cell of the metafilm has dimensions of 
$18\,\U{mm} \times 18\,\U{mm} \times 18\,\U{mm}$.
The conductivity of the metal is 
taken to be $5.8 \times 10^7 \,\U{S/m}$. }
\label{fig:sim}
\end{center}
\end{figure}

The reflection and transmission characteristics of the
metafilm are numerically analyzed using an FDTD 
method to demonstrate the Brewster effect in the
metafilm.  
The simulation system is sketched in
Fig.\,\ref{fig:sim}(a). 
The simulation space has dimensions of 
$2\,\U{m} \times 2\,\U{m} \times 18 \,\U{mm}$ and is discretized into
uniform cubes with dimensions of 
$1\,\U{mm} \times 1\,\U{mm} \times 1\,\U{mm}$. 
Perfectly matched layer (PML) boundary conditions are applied in the $x$
and $y$ directions, while periodic boundary conditions are applied in the
$z$ direction. 
The bi-anisotropic SRRs shown in Fig.\,\ref{fig:sim}(b) are periodically
distributed at $\phi = 45\degree$ to compose the metafilm. 
A Gaussian beam with a spot width of $40\,\U{cm}$ and a Rayleigh
range of $1.3\,\U{m}$ is incident from the left onto the metafilm. 

\begin{figure}[tb]
\begin{center}
\includegraphics[scale=0.7]{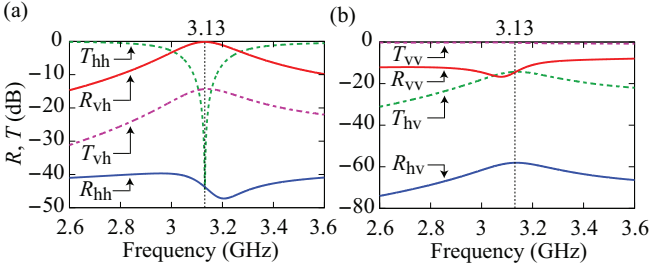}
\caption{Reflection and transmission spectra of the metafilm for 
(a) horizontal and (b) vertical incident polarization.}
\label{fig:spec}
\end{center}
\end{figure}

Figure \ref{fig:spec} plots the reflection and transmission spectra of
the metafilm when an electromagnetic wave is incident from $\phi = 180\degree$. 
The first subscript indicates the
polarization of the reflected/transmitted wave, while the second subscript
indicates the polarization of the incident wave, with v and h referring to vertical and horizontal polarizations,
respectively. 
The reflectance (transmittance) is calculated from the
far-field pattern 
in the direction $\phi = 90\degree$ ($\phi = 0\degree$) 
of the reflected (transmitted) wave. 

Figure \ref{fig:spec}(a) graphs the spectra for horizontal incident
polarization. 
At the resonant frequency of the SRR, namely $3.13\,\U{GHz}$, 
$R\sub{vh}$ has a maximum value and 
$T\sub{hh}$ has a minimum value, for the
following reason. 
The reflected wave is composed solely of the radiated wave from the
metafilm. The amplitude of $\vct{m}$ is largest at the resonant
frequency, and thus the reflectance is maximum. 
On the other hand, the transmitted wave is composed of a superposition
of the incident and radiated waves. 
Since the response of $\vct{p}$ is a resonant (Lorentz) type, the
incident wave and the radiation from $\vct{p}$ cancel each other at the
resonant frequency; thus the
transmittance becomes a minimum. 
The other components, $T\sub{vh}$ and $R\sub{hh}$, should vanish in the
theory but actually have nonzero values due to minor responses of the metafilm
such as multiple scattering and a radiation from an electric dipole moment
$\vct{p}^{\prime}$ excited by the electric field in the $z$ direction,
which are not taken into account in the theory. 
(The response of $\vct{p}^{\prime}$ is non-resonant in this frequency
range. The radiation pattern in the $xy$ plane of $\vct{p}^{\prime}$
is isotropic and the radiation is vertically polarized.)
The ratio $R\sub{vh}/R\sub{hh}$ is larger than $10^4$ at the resonant frequency; 
hence the reflected wave is essentially
vertically polarized. 
Note that $R\sub{vh}/R\sub{hh}$ is large even in the near
off-resonant region. 
The polarization of the reflected wave is vertical 
whenever the principal component of the reflected wave is from
$\vct{m}$. 

\begin{figure}[tb]
\begin{center}
\includegraphics[scale=0.74]{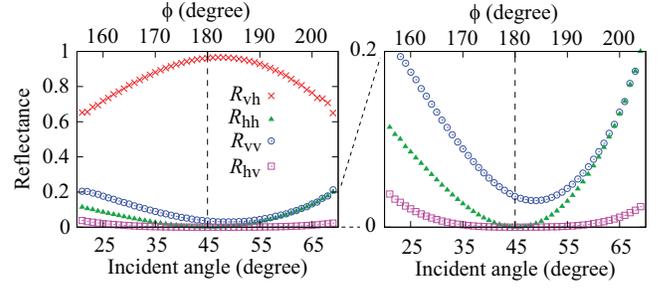}
\caption{Reflectance of the metafilm as a function of the incident angle
for the frequency of $3.13\,\U{GHz}$. The angle $\phi$ is the direction
from which the electromagnetic wave is incident on the metafilm. 
The right panel is a magnification of the left panel in the region of
low reflectance.
}
\label{fig:angle}
\end{center}
\end{figure}

Figure \ref{fig:spec}(b) shows the reflection and transmission spectra for 
vertical incident polarization. 
The interaction between the incident
wave and the metafilm is weak because the vertically polarized incident
wave cannot directly induce $\vct{p}$ and $\vct{m}$. 
Thus $T\sub{vv}$ is nearly 
$0\,\U{dB}$ at every frequency. 
Since a vertically polarized incident wave can directly induce 
$\vct{p}^{\prime}$,
$R\sub{vv}$ is comparable to or larger than the minor components
$T\sub{hv}$ and $R\sub{hv}$. 
The ratio $R\sub{vv}/R\sub{hv}$ is larger than $10^4$ at the resonant
frequency and remains large in the off-resonant region. 
Therefore, the polarization of the reflected wave is also vertical for 
vertical incident polarization. 

Figure \ref{fig:angle} shows the dependence of the reflectance of the
metafilm on the incident angle for the frequency of
$3.13\,\U{GHz}$. $R\sub{hh}$ and $R\sub{hv}$ vanish at the incident
angle of $45\degree$, which reflects the radiation pattern of $\vct{p}$
shown in Fig.\,\ref{fig:theory}(b). On the other hand, $R\sub{vh}$ and
$R\sub{vv}$ have nonzero values at the incident angle of
$45\degree$. This implies that the reflected wave is vertically
polarized regardless of the incident polarization at the incident angle
of $45\degree$. For all angles of incidence except $45\degree$, 
all of the reflectances have nonzero values and such a
polarizing effect does not occur. Therefore, the incident angle of 
$45\degree$ is confirmed to be the Brewster angle of the metafilm.  

If $\vct{p}^{\prime}$ were not induced in the SRR, the Brewster effect
would occur when the plane wave is incident from the direction 
$\phi= 90\degree$. The reflected wave would then become horizontally
polarized. However, that does not occur due to the electric dipole
moment $\vct{p}^{\prime}$ that arises because the SRR
is made of an electric conductor, and a small vertical polarization
component occurs in  the reflected wave for vertical incident polarization. 
If the bi-anisotropic
SRR is instead made of a magnetic conductor, then $\vct{p}^{\prime}$ would vanish and the
Brewster effect would occur for the electromagnetic wave incident from 
$\phi = 90\degree$. 
(Of course, the Brewster effect does not occur for the electromagnetic
wave incident from $\phi = 180\degree$ in this case.)

The Brewster effect in metafilms and that in bulk media have similar
characteristics in the reflection Jones matrix as well as in the
physical meaning. 
The reflection Jones matrix $M\sub{r}$ is defined
by the relation 
$
[E\sub{rv} ~ E\sub{rh}]^t
=
M\sub{r}
[E\sub{iv} ~ E\sub{ih}]^t
=
[\vct{r}_1 ~ \vct{r}_2]^t
[E\sub{iv} ~ E\sub{ih}]^t
$
where $E\sub{r}$ ($E\sub{i}$) is the electric field amplitude of
the reflected (incident)
wave, the second subscript indicates the polarization, 
and $t$ denotes the transpose. 
The column vectors $\vct{r}_1$ and $\vct{r}_2$ that compose $M\sub{r}$
are respectively written as $\vct{r}_1 = [r\sub{vv} ~ r\sub{vh}]^t$ and 
$\vct{r}_2 = [r\sub{hv} ~ r\sub{hh}]^t$ where 
$r_{\alpha \beta} = \sqrt{R_{\alpha \beta}} \exp{[\ii \arg{(r_{\alpha
\beta})}]}$ ($\alpha ,\, \beta = \mathrm{v,\, h}$) is a complex
reflection coefficient.
The Brewster condition in bulk media is given by the 
vanishing of the determinant
of the reflection Jones
matrix~\cite{bassiri88,lakhtakia89,lakhtakia92,tamayama08}.  
For the present metafilm, it is found from Fig.\,\ref{fig:angle} that
$\vct{r}_2$ is zero vector at  the
Brewster angle; therefore, the determinant of the reflection Jones
matrix also vanishes at the Brewster angle. 
Note that the reflected wave vanishes at the Brewster angle for the
incident polarization that satisfies 
$E\sub{iv}/E\sub{ih} = -r\sub{vh}/r\sub{vv}$. 


In conclusion, an extension of the Brewster effect to 
metafilms composed of bi-anisotropic SRRs has been studied
by a theoretical analysis and by full-wave simulations. 
Consideration of the radiation patterns of $\vct{p}$ and
$\vct{m}$ indicates that the polarization of
the reflected wave is independent of the incident polarization when
the SRRs (whose axes are in the direction $\phi = 0\degree$) 
are periodically arranged in the $z$ direction and in the direction 
$\phi = 45\degree$ if the electromagnetic wave is incident on the
metafilm at the incident angle of $45\degree$. 
The reflection and transmission characteristics of the metafilm
have been numerically 
analyzed using an FDTD method. 
The simulation demonstrates that the polarization of the reflected wave
is vertical, independent of the incident
polarization when the electromagnetic wave is incident from the direction
$\phi = 180\degree$ onto the metafilm. 
The determinant of the reflection Jones matrix at the Brewster angle 
has been evaluated 
based on the numerical analysis, and 
the mathematical formulation of 
the Brewster effect in metafilms is found to be the same 
as that in bulk media.
The Brewster effect in the present metafilm can be applied to reflective
optical elements that have combined characteristics of polarizer and
polarization converter. Such Brewster effect has not been observed for
bulk media.  
This study demonstrates that
the Brewster effect can be designed in metafilms by tailoring
a radiation pattern of constituent meta-atoms and a spatial 
distribution of those meta-atoms rather than by varying effective medium
parameters. 
This concept would enable us to create useful optical elements 
that cannot be achieved by the method based on effective medium parameters. 



This study was supported by a Grant-in-Aid for
Scientific Research on Innovative Areas (No. 22109004)
from the Ministry of Education, Culture, Sports, Science,
and Technology of Japan.


\end{document}